\documentclass[a4paper]{article}

\usepackage{INTERSPEECH2022}
\usepackage{cite}

\title{Speak Like a Dog: Human to Non-human creature Voice Conversion}
\name{Kohei Suzuki$^1$,
      Shoki Sakamoto$^1$,
      Tadahiro Taniguchi$^1$
      Hirokazu Kameoka$^2$}
\address{
  $^1$College of Information Science and Engineering, Ritsumeikan University, Japan\\
  $^2$NTT Communication Science Laboratories, NTT Corporation, Japan}
\email{\{suzuki.kohei, sakamoto.shoki, taniguchi\}@em.ci.ritsumei.ac.jp, kameoka.hirokazu@lab.ntt.co.jp}

\begin{document}

\maketitle
\begin{abstract}
  This paper proposes a new voice conversion (VC) task from human speech to dog-like speech while preserving linguistic information as an example of human to non-human creature voice conversion (H2NH-VC) tasks. Although most VC studies deal with human to human VC, H2NH-VC aims to convert human speech into non-human creature-like speech. Non-parallel VC allows us to develop H2NH-VC, because we cannot collect a parallel dataset that non-human creatures speak human language. In this study, we propose to use dogs as an example of a non-human creature target domain and define the ``speak like a dog'' task. To clarify the possibilities and characteristics of the ``speak like a dog'' task, we conducted a comparative experiment using existing representative non-parallel VC methods in acoustic features (Mel-cepstral coefficients and Mel-spectrograms), network architectures (five different kernel-size settings), and training criteria (variational autoencoder (VAE)- based and generative adversarial network-based). Finally, the converted voices were evaluated using mean opinion scores: dog-likeness, sound quality and intelligibility, and character error rate (CER). The experiment showed that the employment of the Mel-spectrogram improved the dog-likeness of the converted speech, while it is challenging to preserve linguistic information. Challenges and limitations of the current VC methods for H2NH-VC are highlighted.
\end{abstract}

\section{Introduction}
\begin{figure}[t]
  \centering
    \includegraphics[width=\linewidth]{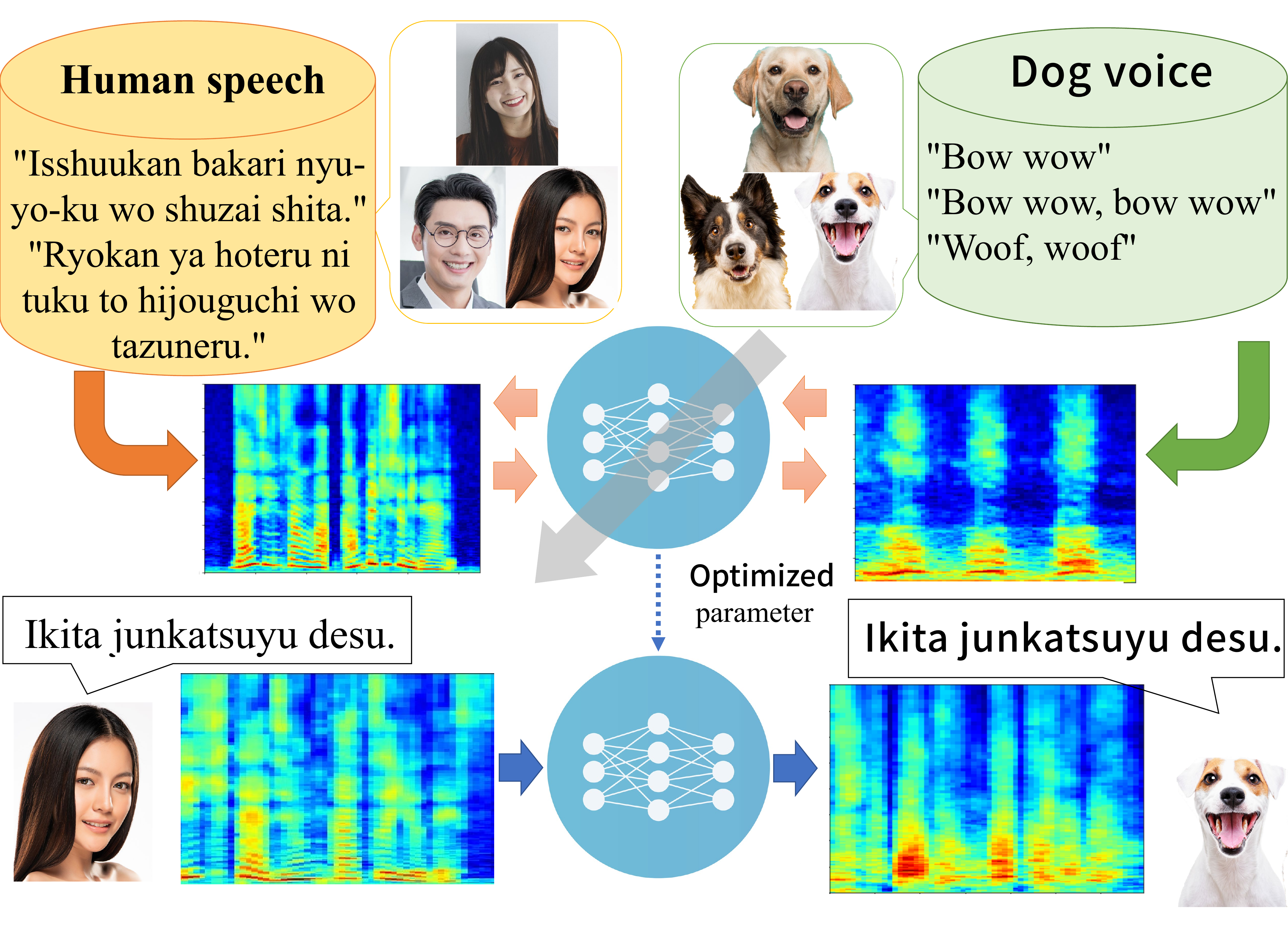}
    \caption{Overview of the human to dog VC. The linguistic information in the training data may differ between speakers.}
    \label{fig:VCprocess}
\end{figure}

Voice conversion (VC) is a technology that converts the speech waveform of the source speaker into a speech waveform with the 
characteristics
%
%
%
of the target speaker while preserving linguistic information~\cite{vc}.
Specifically, in many VC methods, acoustic features are first extracted from the source speaker's speech waveform using speech analysis and then converted to acoustic features similar to those of the target speaker.
Finally, a speech waveform is synthesized using the converted acoustic features.
Most VC studies have focused on human to human VC.
In this study, we consider human to non-human creature VC (H2NH-VC). H2NH-VC converts human voice into non-human creature-like voice while preserving linguistic information.
Non-human creature-like voice refers to voice with certain non-human creature elements such as animals and monsters speaking in a fantasy world.
We expect technologies that can efficiently generate non-human creature-like voice to extend the possibility of creative works in cinema production, game playing, etc.
In this study, we focus on dogs as a representative example of the non-human creature target domain because dog voices are relatively easy to collect and are familiar and prevalent in our daily lives. We define an H2NH-VC task called ``speak like a dog,'' task and investigate the plausibility and challenges of applying existing non-parallel VC methods in this study. 
Figure~\ref{fig:VCprocess} illustrates an overview of the human to dog VC.

Notably, the recent development of non-parallel VC methods~\cite{adain-vc,vqvc,kl,f0vc} has made H2NH-VC a possibility. 
In studies on H2NH-VC, the collection of ground truth speech signals has been a critical barrier to conducting the study. 
Since ``there is no dog in this world that speaks human languages,'' the ground-truth utterance output from the VC system corresponding to the input utterance cannot be obtained.
Therefore, conventional VC methods that require parallel data for training,  parallel VC methods~\cite{dblstm,gmm,dnn}, cannot be used.
For more examples and details of parallel VC methods, readers are referred to a recent review article~\cite{Sisman2020}.

The parallel data consists of source and target speech pairs used to express the same sentence.
In contrast, non-parallel VC methods do not require parallel data. This means that non-parallel VC methods have theoretically overcome the barrier of H2NH-VC.  
Recent deep learning-based, non-parallel VC methods have achieved a VC that is comparable to the ground truth speech in a mean opinion score (MOS) test~\cite{autovc, starganv2}. 
The baseline methods in Voice Conversion Challenge~\cite{vcc2020} are based on generative adversarial networks (GAN) or variational autoencoders (VAE).
One example of a GAN-based method is StarGAN-VC~\cite{sgvc,c-sgvc}, and one example of a VAE-based method is the auxiliary classifier VAE-based VC~(ACVAE-VC)~\cite{acvae}.
However, the applicability of these non-parallel VC methods to H2NH-VC tasks has not been explored. 

To this end, in this paper, we propose a task called ``speak like a dog,''  which is a new task for VC from human voice into dog-like voice while preserving linguistic information. We construct datasets and evaluation criteria for this task.
In addition, we investigated how much VC could be achieved using existing non-parallel VC methods. We experimented with comparing acoustic features, network architectures, and training criteria as baseline methods.

The main contributions of this paper are twofold.
\begin{itemize}
    \item We propose the ``speak like a dog'' task as an example of H2NH-VC tasks and construct a dataset and evaluation criteria.
    \item We investigated the possibilities and characteristics of ``speak like a dog'' task by conducting an experiment comparing existing representative non-parallel VC methods in acoustic features, network architectures, and training criteria. 
\end{itemize}

The remainder of this paper is organized as follows. Section 2 defines the speech task as a dog task. Section 3 describes the methods used in the experiment. Section 4 describes comparative experiments. Finally, Section 5 concludes the paper.



\section{Speak Like a Dog Task}
\subsection{Problem definition}
The proposed ``speak like a dog'' task is one of the H2NH-VC tasks that converts human voice into dog-like voice while preserving linguistic information and representing a dog-like element of the target domain.
Thus, the VC method should preserve linguistic information and represent dog-like elements.
It is not enough to satisfy only one or the other.
In particular, it is important to preserve linguistic information, that is, uttered sentences are recognized correctly by listeners, while ensuring dog-likeness, sound quality, and intelligibility of speech.  

\subsection{Dataset}
A dataset can be non-parallel but should contain human speech signals and dog voices. 
We constructed an example dataset for the ``speak like a dog'' task. Details of the dataset can be found on our website~\footnote{https://github.com/suzuki256/dog-dataset \label{foot}}.
The abstract of the dataset is as follows.

\noindent \textbf{ATR digital sound database.} We use the ATR digital sound database~\cite{atr} for human speech signals.
This is a database of speech recordings of sentences, single words, and other standardized content uttered by two male (MMY and MTK) and two female (FKN and FTK) professional announcers.
The number of each domain's sounds is 503.

\noindent \textbf{Dog dataset.} The dog dataset was constructed from several studies ~\cite{urban,ae,esc}, Freesound project~\cite{freesound}, freesoundslibrary, and Youtube.
Because there is no dog dataset created for VC, we removed extremely soft, loud, and noisy sounds from our collected data.
We also divided them into two datasets based on pitch, adult dog and puppy.
The number of adult dog and puppy sounds is 792 and 288, respectively.
The dataset is available under the terms of the Creative Commons Attribution-NonCommercial license.

\subsection{Evaluation criteria}
\noindent \textbf{Mean Opinion Score~(MOS).}
We defined three MOSs for the ``speak like a dog'' task. They can be obtained by asking subjects to rate the following three MOS tests on a scale of $1$ to $5$:
\begin{enumerate}
    \item {\bf Dog-likeness}: How much of the dog-like element is included?
    
    ($1$ indicates completely not dog-like, and $5$ indicates completely dog-like). 
    \item {\bf Sound quality}: How good is the sound quality?
    
    ($1$ indicates completely low quality, and $5$ indicates completely high quality.)
    \item {\bf Clarity}: How intelligibly were you able to hear the spoken utterance given a written text of the content of the spoken utterance?
    
    ($1$ indicates complete vagueness, and $5$ indicates complete intelligibility). 
\end{enumerate}
\noindent \textbf{Character Error Rate~(CER).}
To evaluate how H2NH-VC preserves linguistic information, we use the character error rate~(CER)~, which represents the error rate between the transcribed sentences of converted speech by the listener and the correct sentences.
The CER is defined as 
\begin{eqnarray}
    \textrm{CER}=\frac{D+S+I}{N},
\end{eqnarray}
where $D,$ $S,$ $I$, and $N$ are the number of deletion errors, substitution errors, insertion errors, and characters in the reference, respectively.

\section{Method}
In this section, we introduce the methods for the experiment.
The key elements involved in VC are acoustic features, network architectures, and the methods of learning the conversion model. Each element has multiple options, but it is not clear what combination is appropriate for converting human voice to dog-like voice.

\subsection{Acoustic features}\label{feature}
The main acoustic features transformed by VC are MCC sequences and the mel-spectrogram.
MCC is an acoustic feature that corresponds to the spectral envelope (shape of the human vocal tract), and the mel-spectrogram is an acoustic feature that includes a harmonic structure (close to the raw acoustic spectrum).
Therefore, it is not clear whether MCC is appropriate for representing a dog's spectral envelope. We predict that the mel-spectrogram works better than MCC in the ``speak like a dog'' task. We then compared the results of the experiments.

\subsection{Network architecture}
Network architectures is another element to investigate. 
In recent years, both convolutional neural network(CNN)-based~\cite{cyclegan,blow,ald,cotatron} and recurrent neural network(RNN)-based network architectures~\cite{phonetic,seq2seq,rnn} have been employed in VC studies.
For example, StarGAN-VC is based on three networks, a generator, discriminator, and domain classifier, with CNN architectures.

This study focuses on a CNN-based network architecture that can be easily investigated by changing the kernel size. 
We want to investigate how much time range the VC system needs to capture in the ``speak like a dog'' task.  
In the experiment, we focus on the kernel size $k$ because it is not obvious how wide a range of time dependencies must be captured in acquiring a model for converting human voice to dog-like voice.

Notably, unlike humans, dogs' voice is short.
Therefore, reducing the kernel size of the discriminator and the domain classifier reduces the receptive field of the CNN. That may potentially results in more dog-like voice.
However, increasing the kernel size increases the amount of information the CNN receives and may allow for a conversion that comes closer to human voice, preserving linguistic information.

\subsection{Training criterion}\label{nonpara-vc}
A VAE-based VC method and a GAN-based VC method have been proposed to enable non-parallel VC.
In this study, we used StarGAN-VC and ACVAE-VC because they are known to function in standard VC tasks as benchmark methods and a type of major generative model.

\begin{figure}[t]
  \centering
    \includegraphics[width=\linewidth]{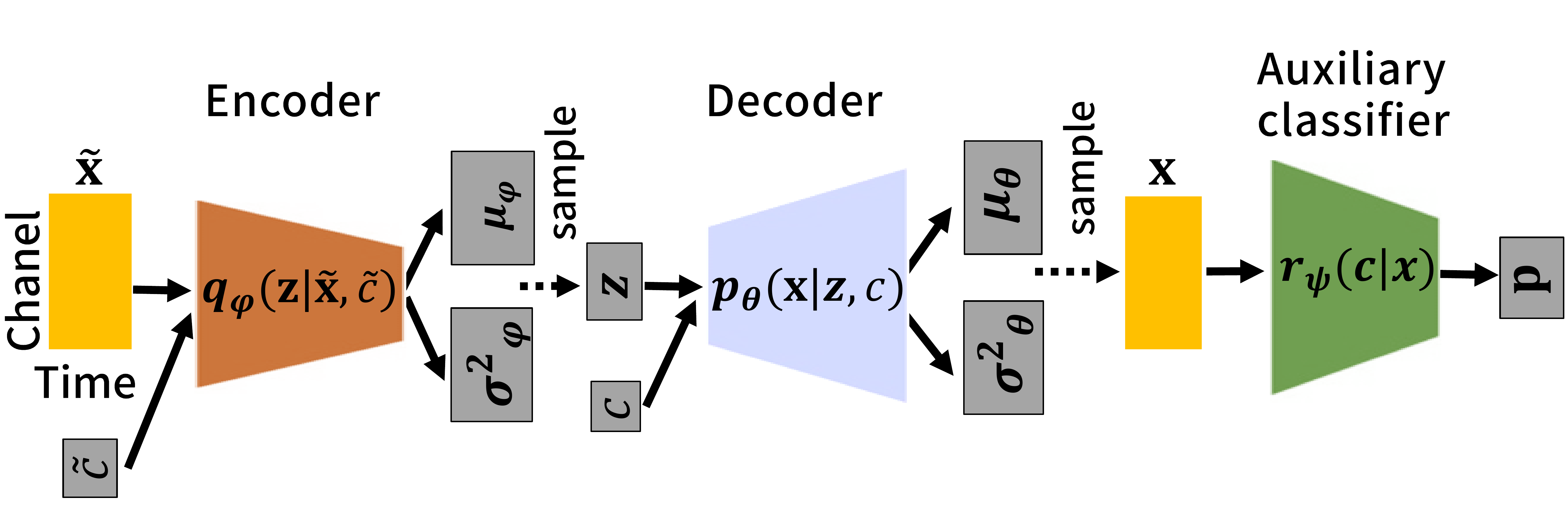}
    \caption{Overview of ACVAE-VC training~\cite{acvae}. $\Tilde{x}$ and $x$ denote acoustic feature sequences.
     $\Tilde{c}$ and $c$ denote speaker information.
     $z$ denote latent variable.
     $q_{\phi}(z|\Tilde{\textbf{x}}, \Tilde{c})$ output the mean $\mu _\phi$ and variance $\sigma _\phi ^2$ of the latent variable $z$ following a normal distribution. $p_{\theta}(\textbf{x}|\textbf{z}, c)$ output the mean $\mu _\theta$ and variance $\sigma _\theta ^2$ of the   \textbf{x} that follows a normal distribution. 
     $r_{\psi}(c | \textbf{x})$ output distribution $p$ which speaker uttered the acoustic feature sequences $x$.}
    \label{fig:acvae}
\end{figure}

\noindent \textbf{ACVAE-VC.}
Figure~\ref{fig:acvae} illustrates an overview of ACVAE-VC.
ACVAE-VC is a VC method that applies the regularization concept of InfoGAN~\cite{infogan} to conditional VAE (CVAE)~\cite{cvae}
%
%
%
%
One of the strengths of this method is its fast learning convergence and stable acquisition of high-performance conversion models.

\begin{figure}[t]
  \centering
    \includegraphics[width=\linewidth]{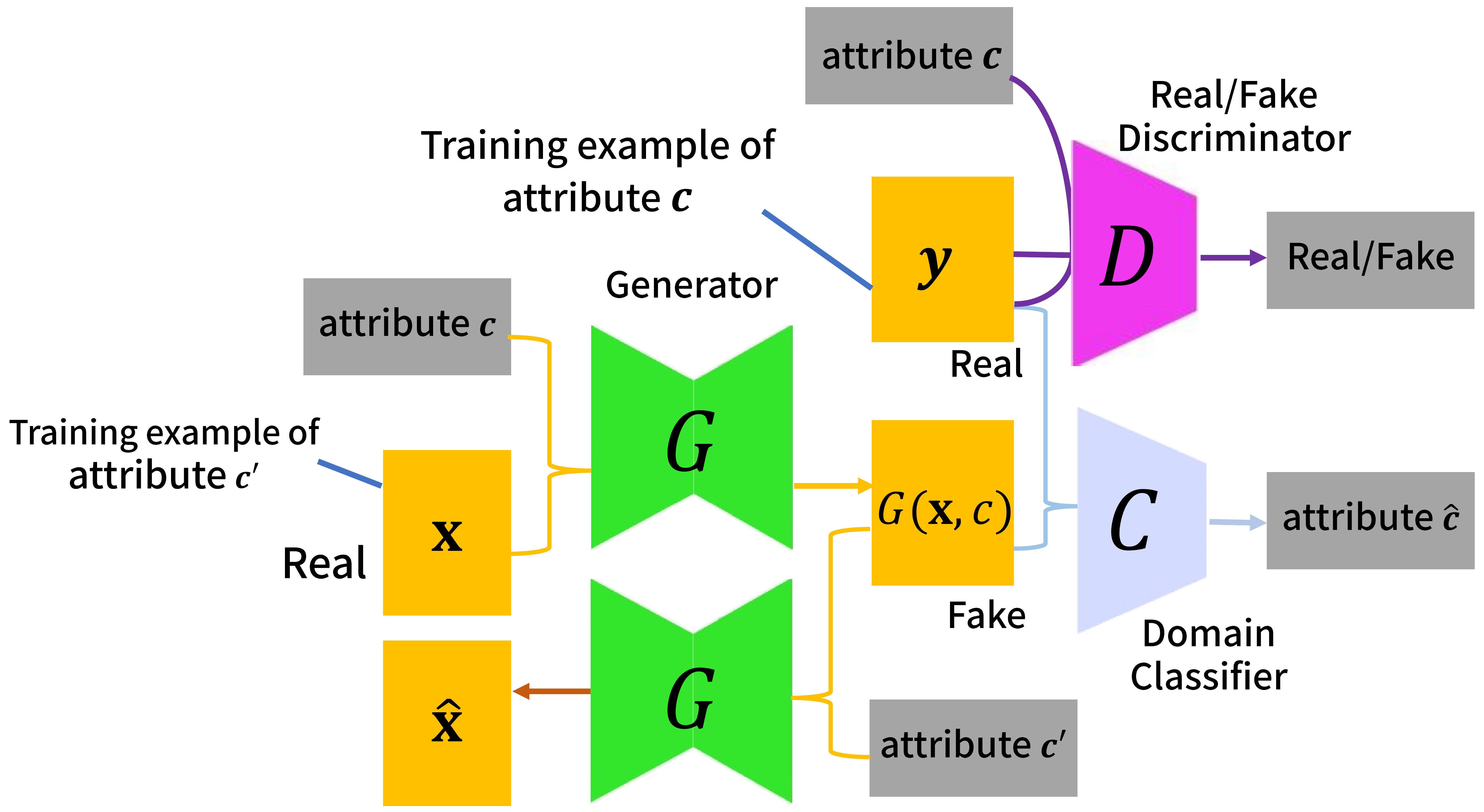}
    \caption{Overview of StarGAN-VC training~\cite{sgvc,c-sgvc}. }
    \label{fig:sgvc}
\end{figure}

\noindent \textbf{StarGAN-VC.}\label{sec:stargan}
Figure~\ref{fig:sgvc} illustrates an overview of StarGAN-VC.
StarGAN-VC is a VC method that can train many-to-many conversion models on non-parallel data based on StarGAN’s ~\cite{stargan} learning method.
One of the strengths of this method is the possibility of obtaining a conversion model that performs as well as or better than ACVAE-VC, if hyperparameters that allow for good learning convergence can be found.

\section{Experiment}
\subsection{Conditions}
The speaker information used for training the VC models were (``FKN,'' ``MMY,'' ``people (FKN, FTK, MMY, and MTK),'' ``adult dog,'' ``puppy,'' and ``dogs (adult\_dog and puppy)'').
Ten recording samples were randomly selected from each domain as evaluation data. The remaining data of each domain were used as the training data.
The sounds used in the evaluation is sounds that converted from FKN to adult\_dog, original sounds of FKN and adult\_dog before the conversion, and white noise.
In this experiment, we generated a speech waveform using WORLD~\cite{world}~(D4C edition~\cite{d4c}) and parallel WaveGAN~\cite{parallel}, and the acoustic features input to the VC model were the MCC sequences and mel-spectrogram, respectively.
We trained Parallel WaveGAN with the same dataset as the VC method.

\noindent \textbf{Experiment 1: comparing acoustic features and learning methods.}
~We performed VC using StarGAN-VC with MCC and the mel-spectrogram and ACVAE-VC with MCC and the mel-spectrogram using default kernel sizes in original paper.

\begin{figure}[t]
  \centering
    \includegraphics[width=\linewidth]{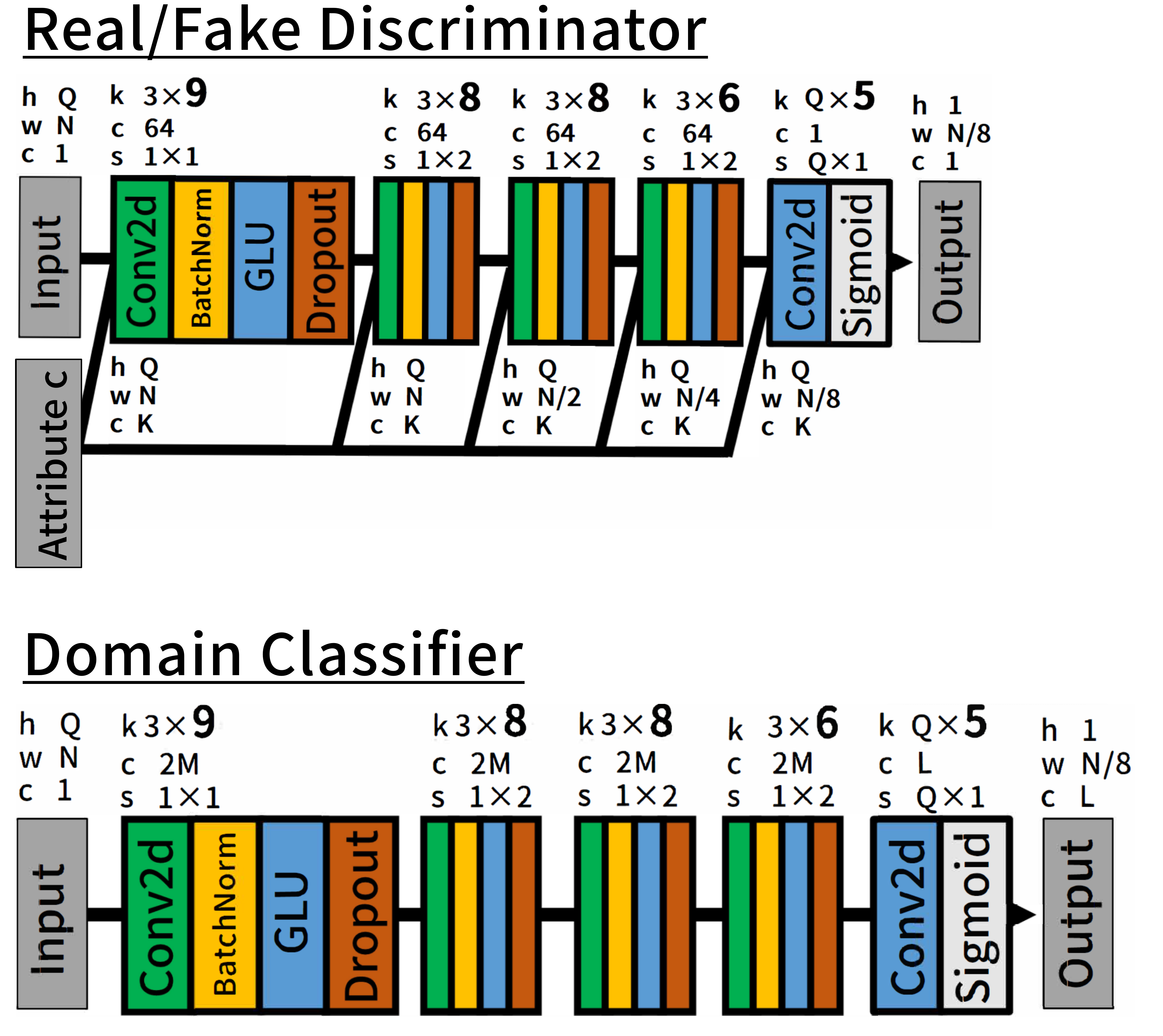}
    \caption{Network architecture of the discriminator and classifier~\cite{stargan,c-sgvc}. In input and output, ``h'', ``w'', and ``c'' represent height, width, and number of channels, respectively. In addition, ``k'', ``c'', and ``s'' denote kernel size, number of channels, and stride size, respectively. Conv1d, BatchNorm, GLU, and Deconv1d denote the 1D convolution layers, batch normalization, gated linear units, and 1D transposed convolution layers, respectively. Batch normalization is applied to each input channel. The class index vectors were repeated along the time direction and then concatenated to the input of each convolution layer.}
    \label{fig:sgvc-network}
\end{figure}

\noindent \textbf{Experiment 2: comparing the kernel size $k$ of the conversion model (CNN).}\label{kernel}
~We define the kernel size $k_d$ of the discriminator and domain classifier described in the original paper as default values. The kernel size  $k_d-2$, $k_d-1$, $k_d$, $k_d+1$, and $k_d+2$ 
in the time direction were compared.
Figure~\ref{fig:sgvc-network} illustrate the original paper's architectures for the discriminator and domain classifier, respectively.
Bold letters indicate the default kernel size $k_d$ for each layer. The value $k_d$ was changed in this experiment.

\subsection{Result}
Tables~\ref{afm1} and \ref{cer1} show the results of the MOS and CER tests in Experiment 1.
In all Tables, ``melspec'' and ``original'' denote the mel-spectrogram and the unconverted test data, respectively.
The bold and underlined numbers in all the tables also denote the highest numerical value of conversion human voice to dog-like voice and the value of voice expected to obtain the highest numerical value in each MOS evaluation, respectively.
The notation inside parentheses next to each VC method in Table \ref{afm1} denotes the input acoustic features.
\begin{table}[tb]
  \caption{Result of the MOS test in Experiment 1}
  \label{afm1}
  \centering
  \begin{tabular}{|c||c|c|c|}  \hline
    Methods & Dog-likeness & \shortstack{Sound\\ quality} & Clarity \\ \hline \hline
    StarGAN-VC~(MCC) & 1.20 & 1.28 & 0.92\\ \hline
    StarGAN-VC~(melspec) & 4.20 & \textbf{2.76} & \textbf{2.04} \\ \hline
    ACVAE-VC~(MCC) & 2.04 & 2.24 & 1.76 \\ \hline
    ACVAE-VC~(melspec) & \textbf{4.24} & 2.36 & 1.36 \\ \hline\hline
    FKN~(original) & 1.00 & \underline{4.80} & \underline{5.00} \\ \hline
    Adult Dog~(original) & \underline{5.00} & 3.70 & 1.00 \\ \hline
    White Noise & 1.10 & 1.20 & 1.00 \\ \hline
  \end{tabular}
\end{table}
\begin{table}[tb]
  \caption{Result of the CER test in Experiment 1}
  \label{cer1}
  \centering
  \begin{tabular}{|c||c|c|c|}  \hline
    Methods & 1st sound & 2nd sound \\ \hline \hline
    StarGAN-VC~(MCC) & 1.00 & 1.00\\ \hline
    StarGAN-VC~(melspec) & \textbf{0.97} & 0.95\\ \hline
    ACVAE-VC~(MCC) & \textbf{0.97} & \textbf{0.94}\\ \hline
    ACVAE-VC~(melspec) & 0.98 & 0.97 \\ \hline\hline
    FKN~(original) & \underline{0.03} & \underline{0.02} \\ \hline
  \end{tabular}
\end{table}

Table \ref{afm1} shows that the method using the mel-spectrogram produces a more dog-like voice than the method using the MCC.
Regarding sound quality, StarGAN-VC with a mel-spectrogram produced the best results.
It is clear that StarGAN-VC with a mel-spectrogram also has the highest clarity of spoken utterances.
In contrast, for ACVAE-VC, we observed that the use of MCC as an acoustic feature synthesizes clearer voice than the use of a mel-spectrogram.

Table \ref{cer1} indicates that there is no significant improvement in CER while the ACVAE-VC with MCC has the smallest error. This shows that the non-parallel VC methods did not successfully achieve the ``speak like a dog'' task preserving linguistic information satisfactorily. However, we can notice some language-like sentences were made from converted voice when listening to the sound generated by ACVAE-VC (MCC). That did not contribute to CER much. 

Table~\ref{tbl:kernel} and \ref{cer2} show the results of the  MOS and CER test in Experiment 2.
\begin{table}[tb]
  \caption{Result of the MOS test in experiment 2}
  \label{tbl:kernel}
  \centering
  \begin{tabular}{|c||c|c|c|}  \hline
    Kernel size & Dog-likeness & \shortstack{Sound\\ quality}  & Clarity \\ \hline \hline
    $k_d$ +2 & 2.20 & 2.40 & 2.00\\ \hline
    $k_d$ +1 & 2.40 & 2.08 & 2.12 \\ \hline
    $k_d$ & \textbf{2.60} & \textbf{2.80} & 2.60 \\ \hline
    $k_d$ -1 & 2.28 & 2.28 & \textbf{3.00} \\ \hline
    $k_d$ -2 & \textbf{2.60} & 2.48 & 2.88 \\ \hline\hline
    FKN~(original) & 1.00 & \underline{4.70} & \underline{5.00} \\ \hline
    Adult Dog~(original) & \underline{5.00} & 3.20 & 1.00 \\ \hline
    White Noise & 1.10 & 1.00 & 1.00 \\ \hline
  \end{tabular}
\end{table}
\begin{table}[tb]
  \caption{Result of the CER in Experiment 2}
  \label{cer2}
  \centering
  \begin{tabular}{|c||c|c|c|}  \hline
    Kernel size & 1st sound & 2nd sound \\ \hline \hline
    $k_d$ +2 & 0.93 & 0.89\\ \hline
    $k_d$ +1 & 0.95 & 0.92\\ \hline
    $k_d$ & 0.97 & 0.95\\ \hline
    $k_d$ -1 & \textbf{0.83} & 0.80 \\ \hline
    $k_d$ -2 & 0.87 & \textbf{0.76} \\ \hline\hline
    FKN~(original) & \underline{0.03} & \underline{0.02} \\ \hline
  \end{tabular}
\end{table}
From Table~\ref{tbl:kernel}, it is clear that the kernel size of the default value and the kernel size of the default value minus $2$ produce a relatively better dog-like voice.
Regarding sound quality, we find that the default values are the best.
It can be seen that the clarity of the spoken utterance is best at the kernel size of the default value minus $1$.
Table \ref{cer2} indicates that the kernel size of the default value $-1$or $-2$ has the smallest error.

\subsection{Discussion}
In the results of the MOS test in Experiment 1, the mel-spectrogram may have a better representation of dog-like voice than the MCC, as predicted by Section~\ref{feature}.

There was no significant difference for dog-like elements between StarGAN-VC and ACVAE-VC using the mel-spectrogram in quantitative evaluation. However, when we listened to the converted sounds, we found qualitative differences subjectively. Sounds produced by ACVAE-VC had a tendency to preserve language-like expressions. The converted sounds can be listened to on our website$^1$.

From the MOS values for clarity in Table~\ref{afm1} and the CER values in Table~\ref{cer1}, ACVAE-VC with the mel-spectrogram preserves less clarity and linguistic information than ACVAE-VC with the MCC.
This may be because the source filter model on which the vocoder is based imitates the human voice production process.

In the results of the CER test of Experiment 1, the high value of CER with existing methods indicates the difficulty of preserving linguistic information in the ``speak like a dog'' task.

In Experiment 2, unlike what was predicted in Section~\ref{kernel}, increasing the kernel size from the default value did not improve the preservation of linguistic information.
This may be because the discriminator and the domain classifier were trained to include the dog's voice and silent intervals by increasing the kernel size.
We found that relatively small kernel size, e.g., $k_d-1$ or $k_d-2$, had a better performance in the ``speak like a dog'' task, generally.

\section{Conclusions}
In this study, we proposed the ``speak like a dog'' task as an example of a human to non-human creature VC task.
Although we could convert human voices into dog-like voices in a fragmented manner, we found that it is challenging to preserve linguistic information.
It was clear that VC methods that worked well for standard VC tasks did not work sufficiently well here.
We also found that using a mel-spectrogram instead of an MCC is important in this task. 


\section{Acknowledgements}
This study was supported by the Japan Society for the Promotion of Science (JSPS) KAKENHI Grant-in-Aid for Scientific Research (A), grant number 21H04904, and JST CREST Grant JPMJCR19A3.

\clearpage
\bibliographystyle{IEEEtran}

\bibliography{main}

\begin{thebibliography}{10}
\providecommand{\url}[1]{#1}
\csname url@samestyle\endcsname
\providecommand{\newblock}{\relax}
\providecommand{\bibinfo}[2]{#2}
\providecommand{\BIBentrySTDinterwordspacing}{\spaceskip=0pt\relax}
\providecommand{\BIBentryALTinterwordstretchfactor}{4}
\providecommand{\BIBentryALTinterwordspacing}{\spaceskip=\fontdimen2\font plus
\BIBentryALTinterwordstretchfactor\fontdimen3\font minus
  \fontdimen4\font\relax}
\providecommand{\BIBforeignlanguage}[2]{{%
\expandafter\ifx\csname l@#1\endcsname\relax
\typeout{** WARNING: IEEEtran.bst: No hyphenation pattern has been}%
\typeout{** loaded for the language `#1'. Using the pattern for}%
\typeout{** the default language instead.}%
\else
\language=\csname l@#1\endcsname
\fi
#2}}
\providecommand{\BIBdecl}{\relax}
\BIBdecl

\bibitem{vc}
Y.~Stylianou, O.~Cappe, and E.~Moulines, ``{{Continuous probabilistic transform
  for voice conversion}},'' \emph{IEEE Transactions on Speech and Audio
  Processing}, vol.~6, no.~2, pp. 131--142, 1998.

\bibitem{adain-vc}
J.-c. Chou, C.-c. Yeh, and H.-y. Lee, ``One-shot voice conversion by separating
  speaker and content representations with instance normalization,''
  \emph{INTERSPEECH}, 2019.

\bibitem{vqvc}
S.~Ding and R.~Gutierrez-Osuna, ``{{Group Latent Embedding for Vector Quantized
  Variational Autoencoder in Non-Parallel Voice Conversion.}}''
  \emph{INTERSPEECH}, pp. 724--728, 2019.

\bibitem{kl}
F.-L. Xie, F.~K. Soong, and H.~Li, ``A kl divergence and dnn-based approach to
  voice conversion without parallel training sentences,'' in
  \emph{INTERSPEECH}, 2016.

\bibitem{f0vc}
\BIBentryALTinterwordspacing
K.~Qian, Z.~Jin, M.~Hasegawa-Johnson, and G.~J. Mysore, ``F0-consistent
  many-to-many non-parallel voice conversion via conditional autoencoder,''
  \emph{IEEE International Conference on Acoustics, Speech and Signal
  Processing (ICASSP)}, May 2020. [Online]. Available:
  \url{http://dx.doi.org/10.1109/ICASSP40776.2020.9054734}
\BIBentrySTDinterwordspacing

\bibitem{dblstm}
L.~Sun, S.~Kang, K.~Li, and H.~Meng, ``Voice conversion using deep
  bidirectional long short-term memory based recurrent neural networks,'' in
  \emph{IEEE International Conference on Acoustics, Speech and Signal
  Processing (ICASSP)}, 04 2015.

\bibitem{gmm}
T.~Toda, A.~W. Black, and K.~Tokuda, ``Voice conversion based on
  maximum-likelihood estimation of spectral parameter trajectory,'' \emph{IEEE
  Transactions on Audio, Speech, and Language Processing}, vol.~15, no.~8, pp.
  2222--2235, 2007.

\bibitem{dnn}
T.~Nakashika, R.~Takashima, T.~Takiguchi, and Y.~Ariki, ``Voice conversion in
  high-order eigen space using deep belief nets,'' in \emph{INTERSPEECH}, 2013.

\bibitem{Sisman2020}
\BIBentryALTinterwordspacing
B.~Sisman, J.~Yamagishi, S.~King, and H.~Li, ``An overview of voice conversion
  and its challenges: From statistical modeling to deep learning,''
  \emph{IEEE/ACM Trans. Audio, Speech and Lang. Proc.}, vol.~29, p. 132–157,
  jan 2021. [Online]. Available:
  \url{https://doi.org/10.1109/TASLP.2020.3038524}
\BIBentrySTDinterwordspacing

\bibitem{autovc}
\BIBentryALTinterwordspacing
K.~Qian, Y.~Zhang, S.~Chang, X.~Yang, and M.~Hasegawa-Johnson, ``{A}uto{VC}:
  Zero-shot voice style transfer with only autoencoder loss,'' in
  \emph{Proceedings of the 36th International Conference on Machine Learning},
  ser. Proceedings of Machine Learning Research, K.~Chaudhuri and
  R.~Salakhutdinov, Eds., vol.~97.\hskip 1em plus 0.5em minus 0.4em\relax PMLR,
  09--15 Jun 2019, pp. 5210--5219. [Online]. Available:
  \url{https://proceedings.mlr.press/v97/qian19c.html}
\BIBentrySTDinterwordspacing

\bibitem{starganv2}
Y.~A. Li, A.~Zare, and N.~Mesgarani, ``{{StarGANv2-VC: A Diverse, Unsupervised,
  Non-parallel Framework for Natural-Sounding Voice Conversion}},''
  \emph{INTERSPEECH}, 2021.

\bibitem{vcc2020}
\BIBentryALTinterwordspacing
Z.~Yi, W.-C. Huang, X.~Tian, J.~Yamagishi, R.~K. Das, T.~Kinnunen, Z.-H. Ling,
  and T.~Toda, ``{Voice Conversion Challenge 2020 –- Intra-lingual
  semi-parallel and cross-lingual voice conversion –-},'' in \emph{Proc.
  Joint Workshop for the Blizzard Challenge and Voice Conversion Challenge
  2020}, 2020, pp. 80--98. [Online]. Available:
  \url{http://dx.doi.org/10.21437/VCC\_BC.2020-14}
\BIBentrySTDinterwordspacing

\bibitem{sgvc}
H.~Kameoka, T.~Kaneko, K.~Tanaka, and N.~Hojo, ``{{Stargan-vc: Non-parallel
  many-to-many voice conversion using star generative adversarial networks}},''
  \emph{IEEE Spoken Language Technology Workshop (SLT)}, pp. 266--273, 2018.

\bibitem{c-sgvc}
H.~Kameoka, T.~Kaneko, K.~Tanaka, and N.~Hojo, ``{{Nonparallel voice conversion
  with augmented classifier star generative adversarial networks}},''
  \emph{IEEE/ACM Transactions on Audio, Speech, and Language Processing},
  vol.~28, pp. 2982--2995, 2020.

\bibitem{acvae}
H.~Kameoka, T.~Kaneko, K.~Tanaka, and N.~Hojo, ``{{ACVAE-VC: Non-parallel voice
  conversion with auxiliary classifier variational autoencoder}},''
  \emph{IEEE/ACM Transactions on Audio, Speech, and Language Processing},
  vol.~27, no.~9, pp. 1432--1443, 2019.

\bibitem{atr}
A.~Kurematsu, K.~Takeda, Y.~Sagisaka, S.~Katagiri, H.~Kuwabara, and K.~Shikano,
  ``Atr japanese speech database as a tool of speech recognition and
  synthesis,'' \emph{Speech Communication}, vol.~9, no.~4, pp. 357--363, 1990.

\bibitem{urban}
\BIBentryALTinterwordspacing
J.~Salamon, C.~Jacoby, and J.~P. Bello, ``A dataset and taxonomy for urban
  sound research,'' in \emph{Proceedings of the 22nd ACM International
  Conference on Multimedia}, ser. MM '14.\hskip 1em plus 0.5em minus
  0.4em\relax New York, NY, USA: Association for Computing Machinery, 2014, p.
  1041–1044. [Online]. Available:
  \url{https://doi.org/10.1145/2647868.2655045}
\BIBentrySTDinterwordspacing

\bibitem{ae}
N.~Takahashi, M.~Gygli, B.~Pfister, and L.~Van~Gool, ``Deep convolutional
  neural networks and data augmentation for acoustic event recognition,'' in
  \emph{INTERSPEECH}, 09 2016, pp. 2982--2986.

\bibitem{esc}
\BIBentryALTinterwordspacing
K.~J. Piczak, ``Esc: Dataset for environmental sound classification,'' in
  \emph{Proceedings of the 23rd ACM International Conference on Multimedia},
  ser. MM '15.\hskip 1em plus 0.5em minus 0.4em\relax New York, NY, USA:
  Association for Computing Machinery, 2015, p. 1015–1018. [Online].
  Available: \url{https://doi.org/10.1145/2733373.2806390}
\BIBentrySTDinterwordspacing

\bibitem{freesound}
\BIBentryALTinterwordspacing
F.~Font, G.~Roma, and X.~Serra, ``Freesound technical demo,'' in
  \emph{Proceedings of the 21st ACM International Conference on Multimedia},
  ser. MM '13.\hskip 1em plus 0.5em minus 0.4em\relax New York, NY, USA:
  Association for Computing Machinery, 2013, p. 411–412. [Online]. Available:
  \url{https://doi.org/10.1145/2502081.2502245}
\BIBentrySTDinterwordspacing

\bibitem{cyclegan}
T.~Kaneko and H.~Kameoka, ``Parallel-data-free voice conversion using
  cycle-consistent adversarial networks,'' \emph{2018 26th European Signal
  Processing Conference (EUSIPCO)}, p. 2100–2104, 2018.

\bibitem{blow}
J.~Serrà, S.~Pascual, and C.~Segura, ``Blow: a single-scale hyperconditioned
  flow for non-parallel raw-audio voice conversion,'' \emph{Advances in Neural
  Information Processing Systems}, p. 6793–6803, 2019.

\bibitem{ald}
J.-c. Chou, C.-c. Yeh, H.-y. Lee, and L.-S. Lee, ``Multi-target voice
  conversion without parallel data by adversarially learning disentangled audio
  representations,'' in \emph{INTERSPEECH}, 09 2018, pp. 501--505.

\bibitem{cotatron}
S.~won Park, D.~young Kim, and M.~chul Joe, ``Cotatron: Transcription-guided
  speech encoder for any-to-many voice conversion without parallel data,'' in
  \emph{INTERSPEECH}, 2020.

\bibitem{phonetic}
L.~Sun, K.~Li, H.~Wang, S.~Kang, and H.~Meng, ``Phonetic posteriorgrams for
  many-to-one voice conversion without parallel data training,'' in \emph{2016
  IEEE International Conference on Multimedia and Expo (ICME)}, 2016, pp. 1--6.

\bibitem{seq2seq}
\BIBentryALTinterwordspacing
J.-X. Zhang, Z.-H. Ling, and L.-R. Dai, ``Non-parallel sequence-to-sequence
  voice conversion with disentangled linguistic and speaker representations,''
  \emph{IEEE/ACM Trans. Audio, Speech and Lang. Proc.}, vol.~28, p. 540–552,
  jan 2020. [Online]. Available:
  \url{https://doi.org/10.1109/TASLP.2019.2960721}
\BIBentrySTDinterwordspacing

\bibitem{rnn}
Y.-A. Chung, W.-N. Hsu, H.~Tang, and J.~Glass, ``{An Unsupervised
  Autoregressive Model for Speech Representation Learning},'' in
  \emph{INTERSPEECH}, 2019, pp. 146--150.

\bibitem{infogan}
X.~Chen, Y.~Duan, R.~Houthooft, J.~Schulman, I.~Sutskever, and P.~Abbeel,
  ``{{Infogan: Interpretable representation learning by information maximizing
  generative adversarial nets}},'' \emph{Proceedings of the 30th International
  Conference on Neural Information Processing Systems}, pp. 2180--2188, 2016.

\bibitem{cvae}
D.~P. Kingma, S.~Mohamed, D.~Jimenez~Rezende, and M.~Welling,
  ``{{Semi-supervised Learning with Deep Generative Models}},'' \emph{Advances
  in Neural Information Processing Systems}, vol.~27, 2014.

\bibitem{stargan}
Y.~Choi, M.~Choi, M.~Kim, J.-W. Ha, S.~Kim, and J.~Choo, ``{{Stargan: Unified
  generative adversarial networks for multi-domain image-to-image
  translation}},'' \emph{Proceedings of the IEEE conference on computer vision
  and pattern recognition}, pp. 8789--8797, 2018.

\bibitem{world}
M.~Morise, F.~Yokomori, and K.~Ozawa, ``{{WORLD: a vocoder-based high-quality
  speech synthesis system for real-time applications}},'' \emph{IEICE
  TRANSACTIONS on Information and Systems}, vol.~99, no.~7, pp. 1877--1884,
  2016.

\bibitem{d4c}
M.~Morise, ``{{D4C, a band-aperiodicity estimator for high-quality speech
  synthesis}},'' \emph{Speech Communication}, vol.~84, pp. 57--65, 2016.

\bibitem{parallel}
R.~Yamamoto, E.~Song, and J.-M. Kim, ``Parallel wavegan: A fast waveform
  generation model based on generative adversarial networks with
  multi-resolution spectrogram,'' \emph{IEEE International Conference on
  Acoustics, Speech and Signal Processing (ICASSP)}, pp. 6199--6203, 2020.

\end{thebibliography}


\end{document}